# Direct Visualization of Perm-Selective Ion Transportation

Wonseok Kim[1,2], Jungeun Lee[3], Gun Yong Sung[3]* and Sung Jae Kim[1,4,5]*

[1]*Department of Electrical and Computer Engineering,*
*Seoul National University, Seoul, 08826, Republic of Korea*
[2]*Department of Bioengineering,*
*University of California Berkeley, CA 94720, USA*
*Seoul National University, Seoul, 08826, Republic of Korea*
[3]*Department of Material Science and Engineering,*
*Hallym University, Chuncheon, 24252 Republic of Korea*
[4]*Nano System Institute, Seoul National University, Seoul. 08826, South Korea*
[5]*Inter-university Semiconductor Research Center,*
*Seoul National University, Seoul, 08826, South Korea*

*E-mail: (SJKim) gates@snu.ac.kr, Phone: +82-2-880-1665,
(GYSung) gysung@hallym.ac.kr, Phone: +82-33-248-2361.

**ABSTRACT**

Perm-selective ion transportation in a nanoscale structure has been extensively studied with aids of nanofabrication technology for a decade. While theoretical and experimental advances pushed the phenomenon to seminal innovative applications, its basic observation has relied only on an indirect analysis such as current-voltage relation or fluorescent imaging adjacent to the nanostructures. Here we experimentally, for the first time, demonstrated a direct visualization of perm-selective ion transportation through the nanostructures using an ionic plasma generation. A micro/nanofluidic device was employed for a micro bubble formation, plasma negation and penetration of the plasma through the nanojunction. The direct observation provided a keen evidence of perm-selectivity, *i.e.* allowing cationic species and rejecting anionic species. Furthermore, we can capture the plasma of $Li^+$, which has lower mobility than $Na^+$ in aqueous state, passed the nanojunction faster than $Na^+$ due to the absence of hydrated shells around $Li^+$. This simple, but essential visualization technique would be effective means not only for advancing the fundamental nanoscale electrokinetic study but also for providing the insight of new innovative engineering applications.

# Introduction

Ion selective transportation through nanoporous membrane (or nanochannel) has been extensively studied due to its fundamental importance to understand basic biological functions[1-3] and developing innovative engineering applications such as electro-desalination[4-6] and high energy efficient battery[7-9]. Especially, single ion selective pump has been drawn ever-increasing attentions, for example, potassium selective pump and sodium selective pump in cell membrane for physiological homeostasis of living organism[10, 11] or high voltage generation of electrical eel[12]. More recently, extracting lithium ions out of abundant sodium ions in seawater has become key issues in lithium ion battery industry for supplying economic raw material[13, 14]. Such high industrial demands should require a proper understanding of nanostructure and its fundamental interaction with adjacent electrolyte solution.

Engineered nanostructures has a characteristic length scale less than electrical double layer of few nanometers formed adjacent to the interface of electrolyte/solid[15]. Thus, electrical polarity inside a confined nanostructure should be either positive or negative depending on the surface charge of solid substrate. If the solid such as silicon and glass is negatively charged, only cations can pass through the nanostructure, while anions are rejected to enter the nanostructure. This is conventional working principle of perm selectivity[16]. While several works successfully demonstrated the working principles, major nuisance is that their investigations typically have relied on indirect methods such as measuring current-voltage relation[17-19], (fluorescent) imaging[20-22] or analyzing a transported sample fluid by mass spectrometry[23, 24], *etc*. mainly because direct and *in situ* visualization inside nanostructure is highly challenging task. As a result, profound debates are still ongoing for the origin of ion selectivity and its related phenomena that has never been demonstrated. In this sense, the direct visualization of perm-selective ion transportation through nanostructure becomes priceless subject to be investigated.

On the other hand, studies related to the plasma generation at the atmospheric pressure and the low temperature condition in the microfluidic device have been reported[25, 26], but there have been few studies on the utility of inherent luminescence characteristics due to the obvious disadvantages such as high voltage operation and the instability[27, 28]. Here, we employed this plasma discharge for reporting the direct observation of perm selective transportation for the first time.

# Methods

**Device fabrication.** The microchannels were molded by the general PDMS fabrication process[29]. Briefly, PDMS solution at the ratios of pre-polymer (PDMS, Sylgard 184, Dow corning) to curing agent of 10:1 was mixed and degassed for an hour. After pouring on lithographically constructed Si wafer, it was cured at 75°C for 4 hours. The demolded PDMS block and Nafion (Sigma Aldrich, USA)-patterned glass substrate were irreversibly adhered by $O_2$ plasma treatment (Cute-MP, FemtoScience, Korea). The Nafion was patterned by surface patterned method[30, 31]. Both microchannels were filled with 2 M NaCl or 2 M LiCl solution.

**Experimental setups.** Ag/AgCl electrodes were inserted at both reservoirs to apply external voltage (PS 350, Stanford Research Systems, USA or 237 High voltage source measure unit, Keithley, USA). Current value at each voltage step was obtained by the customized Labview program. An inverted fluorescence microscope (IX-51, Olympus, Japan) and a CCD camera (DP73, Olympus, Japan) were used to detect and trace ionic plasma image. Commercial software (CellSense, Olympus, Japan) was used to synchronize the CCD camera with the microscope and to analyze the images. Time evolving snapshots of nanoelectrokinetic lightning were captured using a high-speed camera (Fastcam Mini UX50, Photron, Japan).

## Results and Discussions

**A new concept for direct visualization.** A schematic illustration of direct visualization was provided in Figure 1(a). Applying relatively low voltage on anodic channel, a bubble was generated due to electrical breakdown as shown in step 1 of Figure 1(a). High voltage, then, was applied to the anodic channel to initiate the plasma discharge inside the bubble, since high electrical resistance focused the electric potential only to the bubble (step 2 in Figure 1(a)). Plasma, so-called the fourth fundamental state of matter, can be generated by subjecting a gas to a strong electric field and the gas becomes an ionized gaseous substance. In this experiment, since either NaCl or KCl were injected in the microchannel, $Na^+$, $Li^+$ and $Cl^-$ would be the possible ions turned to be plasma. Finally, the plasma can penetrate through the nanojunction by electrical grounding the cathodic channel (step 3 in Figure 1(a)). In order to demonstrate this concept, a polydimethylsiloxane (PDMS) microchip incorporated with Nafion nanoporous membrane was fabricated as shown in Figure 1(b). PDMS and glass should be the building block of the microchip, because silicon substrate has lower dielectric breakdown voltage than water so that it destroys when high voltage is applied. Simple straight microchannels were parallelly aligned and Nafion nanojunction (2 mm (length) × 1 μm (depth) × 2 mm (width)) was bridging the microchannels to guide plasma transportation. Nafion contains long chained polymer with highly charged functional group so that it acts as perm-selective material because its pore size distributes in the rage of 1 nm – 10 nm[32]. This configuration has been investigated for a couple of decades to develop efficient biomolecular preconcentrator[33-38] or nanoelectrokinetic desalination / purification platform[5, 39, 40]. Both microchannels had the dimension of 2,000 mm (length) × 15 μm (depth) × 1 mm (width). The fabrication of PDMS microchannel followed general soft-lithographical method[29] and Nafion was patterned using surface patterning method[30, 41]. Aforementioned schematics of each step were experimentally demonstrated as shown in Figure 2.

**Micro bubble formation**. Either NaCl (2 M) or LiCl (2 M) was injected into both channel and electrical voltage was applied only through the anodic channel in step 1 using high voltage power supplier (PS 350, Stanford Research Systems, USA or 237 High voltage source measure unit, Keithley, USA) under microscopic observation by inverted microscope (IX53, Olympus, Japan). As the voltage increase, current initially increased linearly but it suddenly dropped after 50 V, indicating an electrical shortage due to a bubble generation (Figure 2(a)). In the meantime, microscopic observation verified the formation of bubble inside the anodic channel. This was attributed to the electric breakdown and Joule heating over a threshold voltage. While the bubble was generated at random position, the location should be inside the anodic channel because the electrical resistance was higher in the microchannel than one in reservoir. Also, only one bubble was formed since the electric field was focused to the bubble once the bubble was generated. See Supporting Video 1 for the micro bubble formation. Note that the electric current was still measurable after forming the bubble mainly because remaining thin liquid film on the surface of PDMS still conducted the current.

**Plasma generation.** Switching the electrical voltage over 500 V enabled to discharge a plasma inside the micro bubble (Figure 2(b)). Since the vapor inside the bubble contained $Cl^-$ in common and either $Li^+$ or $Na^+$, unique flame colors were emitted from the plasma. (Representing color of each ions were violet, red and yellow for $Cl^-$, $Li^+$ and $Na^+$, respectively.[42]) Microscopic images in Figure 2(b) were captured even without background light because of a strong emission from the plasma itself. See Supporting Video 2 for each electrolyte solution. Image analysis (ImageJ, NIH and Photo Shop, Abode) confirmed that the plasma emitted all of three colors inside the bubble. Black background and saturated area were subtracted from original image and counting the number of pixels for three color bands. Then, the values were normalized with the maximum value among them. Note that conventional photo-spectrometer was unable to measure the color spectrum due to fast and fluctuating plasma generation[27, 28]. While red and yellow were mixed in both cases of NaCl and LiCl, the important observation was that the violet, which is unique emission of chlorine plasma, was comparable to other colors.

**Nanoelectrokinetic lightning**. In the meantime, the plasma slowly moved toward Nafion nanojunction and penetrated the nanojunction by electrical grounding the cathodic microchannel in step 3. Since plasma and Nafion are conductive matter, electroosmotic flow pushed the bubble toward the nanojunction[43]. Time evolving snapshots were given in Figure 2(c) using a high-speed camera (Fastcam Mini UX50, Photron, Japan). See Supporting Video 3 for the plasma penetration through nanojunction.

The featured observation in this work was that the color of plasma during the penetration mostly excluded the purple emission from $Cl^-$ in both cases of NaCl (Figure 3(a)) and LiCl (Figure 3(b)). This reflected that anionic species was rejected by cation selective nanojunction. Since plasma retains an ionic gaseous state, plasma of $Cl^-$ still possesses an anionic characteristic. Whereas yellowish and reddish plasma can pass through the nanojunction, we can visually confirm the cation selective ion transportation. Note that 250 of high-speed images were overlaid in Figure 3(a) and 3(b). While clear transportations were captured in the case of NaCl (Figure 3(a)), only few of lightning was observed in LiCl (Figure 3(b)). This was attributed to faster mobility of $Li^+$ than one of $Na^+$. In aqueous state, the mobility of $Li^+$ is smaller than one of $Na^+$ due to hydrated shell of water molecules[43, 44]. However, $Li^+$ can move without such heavy shells in plasma state so that it transported faster than $Na^+$. Thus, only fewer lightning was captured at the same frame rate of high-speed imaging.

This velocity difference was directly observable with the mixture of NaCl (1 M) and LiCl (1 M) solution. Both electrolytes were injected into the anodic microchannel and the same experimental procedures were repeated. As shown in Figure 3(c), the plasma of both $Na^+$ and $Li^+$ with minimum purple emission was passing through the nanojunction. One strip of plasma was magnified in Figure 3(d). As directly visualized here, reddish plasma always leads yellowish plasma. This observation clearly demonstrated faster mobility of $Li^+$ than one of $Na^+$ when there were no hydrated shells.

## Conclusions

Here we experimentally demonstrated a clear evidence of perm-selective ion transportation through nanoporous membrane by the direct visualization using a microfluidic plasma generation. Micro/nanofluidic platform was employed to generate an ionic plasma inside a microchannel. By pushing the plasma into the nanojunction, light emissions only from cationic species penetrated the nanojunction, while one from anionic species was rejected from entering the nanojunction. It, for the first time, visually and directly confirmed the *in situ* perm-selective ion transportation. More importantly, the plasma of $Li^+$, which has lower mobility than $Na^+$ in aqueous state, passed the nanojunction faster than $Na^+$ since hydrated shells around $Li^+$ were stripped out in ionic plasma state. Simple but effective demonstration of this visualization would be an effective mean not only for characterizing the ion-selectivity of nanostructures but also for answering fundamental questions arisen in nanoscale electrokinetic research field. Note that we also have tried with $K^+$, which has greenish emission, but its emission was too low to be detected. In engineering aspect, however, one could utilize this technique to detect unknown cationic species from certain sample fluid with minimum sample volume, if one setup photo-analysis tools of high-resolution.

**Supporting Information**

Video1. Micro bubble formation inside a microchannel (avi)

Video2. Plasma generation inside the micro bubble (avi)

Video3. Nanoelectrokinetic lightning through nanoporous membrane (avi)

**Author Contributions**

Conceptualization: Wonseok Kim

Investigation: Wonseok Kim and Jungeun Lee

Methodology: Wonseok Kim

Supervision: Gun Yong Sung and Sung Jae Kim

Writing: Wonseok Kim, Jungeun Lee, Gun Yong Sung and Sung Jae Kim

**Competing interests**

The authors declare no competing interests (both financial and non-financial).

**Acknowledgements**

This work is supported by Basic Research Laboratory Project (NRF-2018R1A4A1022513) and the Center for Integrated Smart Sensor funded as Global Frontier Project (CISS- 2011-0031870) by the Ministry of Science and ICT. Also all authors acknowledged the financial supports from BK21 Plus program of the Creative Research Engineer Development IT, Seoul National University. S. J. Kim acknowledged the financial support from LG Yonam Foundation, Korea.

**FIGURE CAPTIONS**

**Figure 1** (a) Schematic diagrams of micro/nanofluidic platform for the direct visualization of perm-selective ion transportation using an ionic plasma generation. Experimental steps were (i) micro bubble formation inside a microchannel, (ii) plasma generation in the microchannel and (iii) plasma penetration through a nanojunction. (b) Fabricated micro/nanofluidic device.

**Figure 2** Experimental demonstrations of the visualization. (a) Micro bubble formation with electrical current plot as a function of applied voltage. (b) Plasma generation using high electric voltage for NaCl and KCl electrolyte solution. (c) Time evolving snapshots of the penetration of the plasma through the nanojunction.

**Figure 3** Overlaid images of plasma penetration in the nanojunction for (a) NaCl, (b) KCl and (c) the mixture of NaCl and KCl. (d) Magnified snapshot of one strip of plasma to identify the faster migration of $Li^+$ plasma than that of $Na^+$ plasma.

# REFERENCES

[1] Branton D, Deamer DW, Marziali A, Bayley H, Benner SA, Butler T, et al. The potential and challenges of nanopore sequencing. Nanoscience And Technology: A Collection of Reviews from Nature Journals: World Scientific; 2010. p. 261-8.

[2] Venkatesan BM, Bashir R. Nanopore sensors for nucleic acid analysis. Nat Nanotech. 2011;6:615.

[3] Meller A, Nivon L, Brandin E, Golovchenko J, Branton D. Rapid nanopore discrimination between single polynucleotide molecules. Proc Natl Acad Sci U S A. 2000;97:1079-84.

[4] Nikonenko VV, Kovalenko AV, Urtenov MK, Pismenskaya ND, Han J, Sistat P, et al. Desalination at overlimiting currents: State-of-the-art and perspectives. Desalination. 2014;342:85-106.

[5] Kim SJ, Ko SH, Kang KH, Han J. Direct seawater desalination by ion concentration polarization. Nat Nanotech. 2010;5:297-301.

[6] Knust KN, Hlushkou D, Anand RK, Tallarek U, Crooks RM. Electrochemically Mediated Seawater Desalination. Angewandte chemie International Edition. 2013;52:8107-10.

[7] Choi SW, Jo SM, Lee WS, Kim YR. An electrospun poly (vinylidene fluoride) nanofibrous membrane and its battery applications. Advanced Materials. 2003;15:2027-32.

[8] Jia C, Liu J, Yan C. A significantly improved membrane for vanadium redox flow battery. J Power Sources. 2010;195:4380-3.

[9] Song Y-A, Batista C, Sarpeshkar R, Han J. In-plane integration of ion-selective membrane in microfluidic PEM fuel cell by micro flow surface patterning. J Power Sources. 2008;183:674-7.

[10] Gumz ML, Rabinowitz L, Wingo CS. An integrated view of potassium homeostasis. New Engl J Med. 2015;373:60-72.

[11] Palmer BF. Regulation of potassium homeostasis. Clinical Journal of the American Society of Nephrology. 2015;10:1050-60.

[12] Schroeder TB, Guha A, Lamoureux A, VanRenterghem G, Sept D, Shtein M, et al. An electric-eel-inspired soft power source from stacked hydrogels. Nature. 2017;552:214.

[13] Chung K-S, Lee J-C, Kim W-K, Kim SB, Cho KY. Inorganic adsorbent containing polymeric membrane reservoir for the recovery of lithium from seawater. J Membr Sci. 2008;325:503-8.

[14] Umeno A, Miyai Y, Takagi N, Chitrakar R, Sakane K, Ooi K. Preparation and adsorptive properties of membrane-type adsorbents for lithium recovery from seawater. Industrial & engineering chemistry research. 2002;41:4281-7.

[15] Probstein RF. Physicochemical Hydrodynamics : An Introduction: Wiley-Interscience; 1994.

[16] Schoch RB, Han J, Renaud P. Transport phenomena in nanofluidics. Rev Mod Phys. 2008;80:839-83.

[17] Schiffbauer J, Park S, Yossifon G. Electrical Impedance Spectroscopy of Microchannel-Nanochannel Interface Devices. Phys Rev Lett. 2013;110.

[18] Sohn S, Cho I, Kwon S, Lee H, Kim SJ. Surface Conduction in a Microchannel. Langmuir. 2018;34:7916-21.

[19] Rubinstein I, Zaltzman B. Dynamics of extended space charge in concentration polarization. Phys Rev E. 2010;81:061502

[20] Rubinstein SM, Manukyan G, Staicu A, Rubinstein I, Zaltzman B, Lammertink RGH, et al. Direct Observation of a Nonequilibrium Electro-Osmotic Instability Phys Rev Lett. 2008;101:236101.

[21] Kim SJ, Wang Y-C, Lee JH, Jang H, Han J. Concentration Polarization and Nonlinear Electrokinetic Flow near Nanofluidic Channel. Phys Rev Lett. 2007;99:044501.

[22] Kim K, Kim W, Lee H, Kim SJ. Stabilization of ion concentration polarization layer using micro fin structure for high-throughput applications. Nanoscale. 2017;9:3466-75.

[23] Kim W, Park S, Kim K, Kim SJ. Experimental verification of simultaneous desalting and molecular preconcentration by ion concentration polarization. Lab Chip. 2017;17:3841-50.

[24] Piruska A, Gong M, Sweedler JV, Bohn PW. Nanofluidics in chemical analysis. Chem Soc Rev. 2010;39:1060-72.

[25] Bruggeman P, Degroote J, Vierendeels J, Leys C. DC-excited discharges in vapour bubbles in capillaries. Plasma Sources Science and Technology. 2008;17:025008.

[26] Eijkel JC, Stoeri H, Manz A. A molecular emission detector on a chip employing a direct current microplasma. Anal Chem. 1999;71:2600-6.

[27] Stenzel R. Instability of the sheath-plasma resonance. Phys Rev Lett. 1988;60:704.

[28] Kato Y, Mima K, Miyanaga N, Arinaga S, Kitagawa Y, Nakatsuka M, et al. Random phasing of high-power lasers for uniform target acceleration and plasma-instability suppression. Phys Rev Lett. 1984;53:1057.

[29] Duffy DC, McDonald JC, Schueller OJA, Whitesides GM. Rapid Prototyping of Microfluidic Systems in Poly(dimethylsiloxane). Anal Chem. 1998;70:4974-84.

[30] Cho I, Sung G, Kim SJ. Overlimiting Current Through Ion Concentration Polarization Layer: Hydrodynamic

Convection Effects. Nanoscale. 2014;6:4620-6.
[31] Choi J, Huh K, Moon DJ, Lee H, Son SY, Kim K, et al. Selective preconcentration and online collection of charged molecules using ion concentration polarization. RSC Advances. 2015;5:66178-84.
[32] Mauritz KA, Moore RB. State of understanding of Nafion. Chem Rev. 2004;104:4535-85.
[33] Wang Y-C, Stevens AL, Han J. Million-fold Preconcentration of Proteins and Peptides by Nanofluidic Filter. Anal Chem. 2005;77:4293-9.
[34] Hong SA, Kim Y-J, Kim SJ, Yang S. Electrochemical detection of methylated DNA on a microfluidic chip with nanoelectrokinetic pre-concentration. Biosensors and Bioelectronics. 2018;107:103-10.
[35] Lee H, Choi J, Jeong E, Baek S, Kim HC, Chae J-H, et al. dCas9-mediated Nanoelectrokinetic Direct Detection of Target Gene for Liquid Biopsy. Nano Lett. 2018;18:7642-50.
[36] Fu LM, Hou HH, Chiu PH, Yang RJ. Sample preconcentration from dilute solutions on micro/nanofluidic platforms: A review. Electrophoresis. 2018;39:289-310.
[37] Kim SJ, Song Y-A, Han J. Nanofluidic concentration devices for biomolecules utilizing ion concentration polarization: theory, fabrication, and application. Chem Soc Rev. 2010;39:912-22.
[38] Son SY, Lee S, Lee H, Kim SJ. Engineered nanofluidic preconcentration devices by ion concentration polarization. BioChip Journal. 2016;10:251-61.
[39] Kwak R, Kim SJ, Han J. Continuous-flow biomolecule and cell concentrator by ion concentration polarization. Anal Chem. 2011;83:7348-55.
[40] Kim B, Kwak R, Kwon HJ, Pham VS, Kim M, Al-Anzi B, et al. Purification of High Salinity Brine by Multi-Stage Ion Concentration Polarization Desalination. Sci Rep. 2016;6:31850.
[41] Lee J-H, Song Y-A, Kim SJ, Han J. High-Throughput Proteomic Sample Preconcentration in PDMS Microfluidic Chip Using Surface-Patterned Ion-Selective Membrane. Proceedings of the MicroTAS 2007 Conference. Vol. 2. Paris, France2007. p. 1198-200.
[42]
[43] Kirby BJ. Micro- and Nanoscale Fluid Mechanics: Cambridge University Press; 2010.
[44] Choi W, Ulissi ZW, Shimizu SFE, Bellisario DO, Ellison MD, Strano MS. Diameter-dependent ion transport through the interior of isolated single-walled carbon nanotubes   Nature Communications. 2013;4:2397.

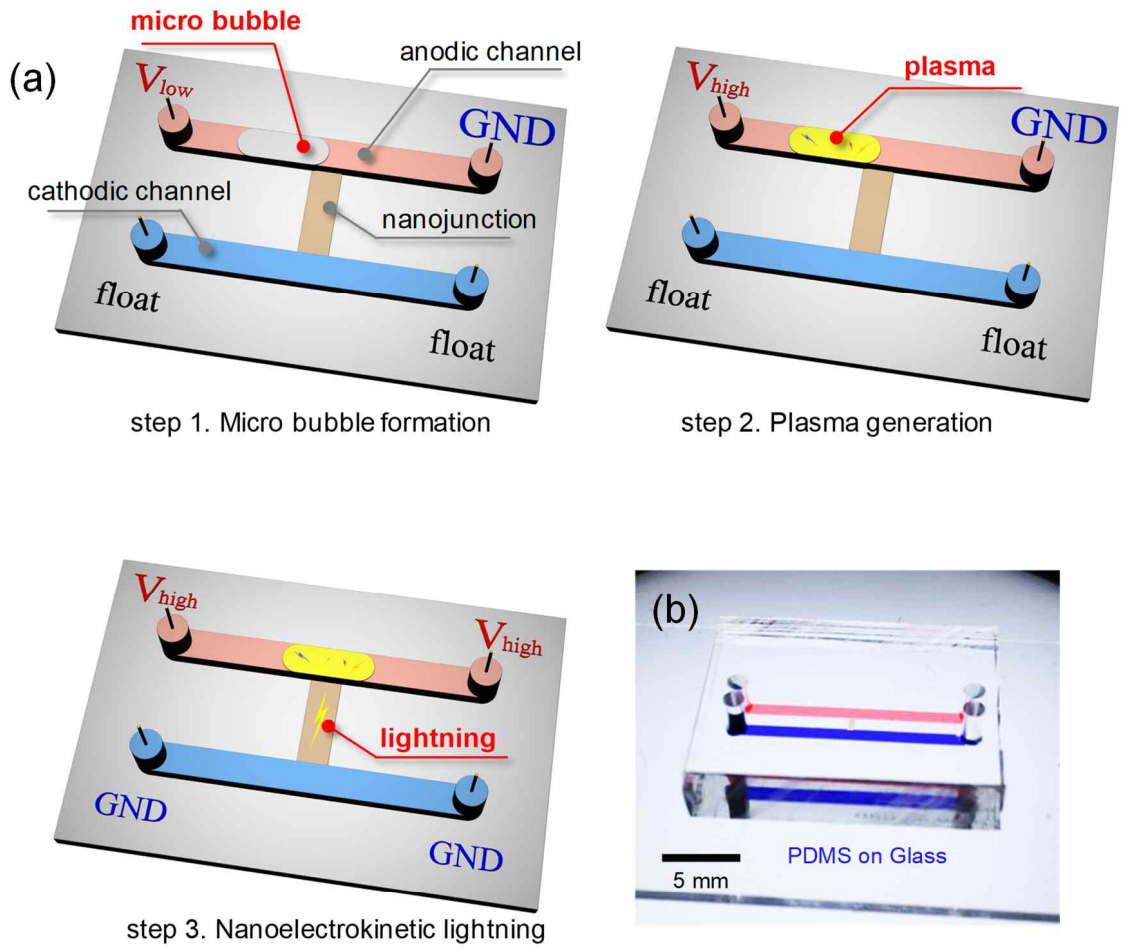

Figure 1

(a) step 1. Micro bubble formation

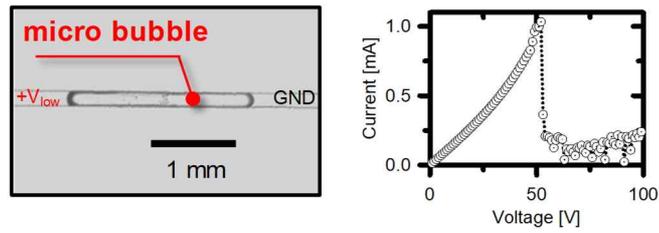

(b) step 2. Plasma generation

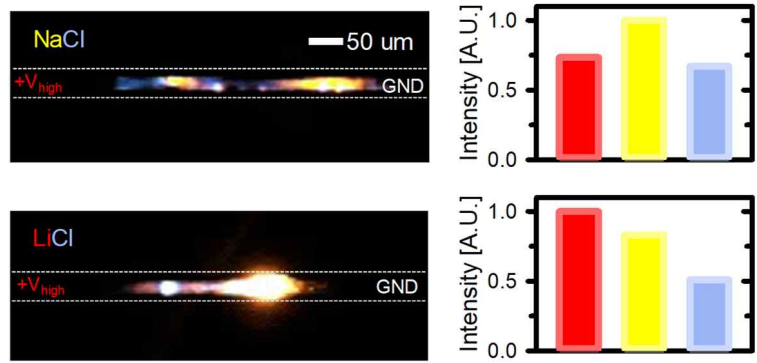

(c) step 3. Nanoelectrokinetic lightning

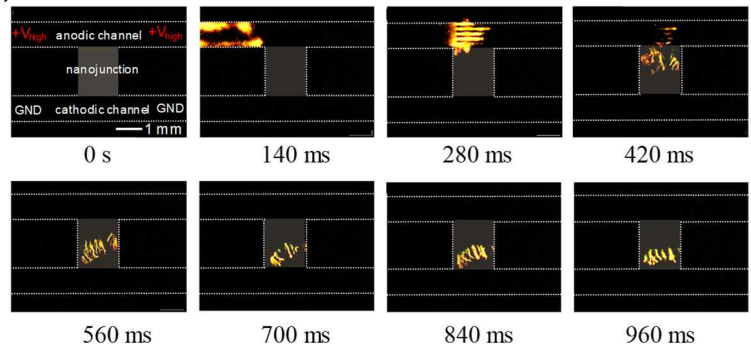

Figure 2

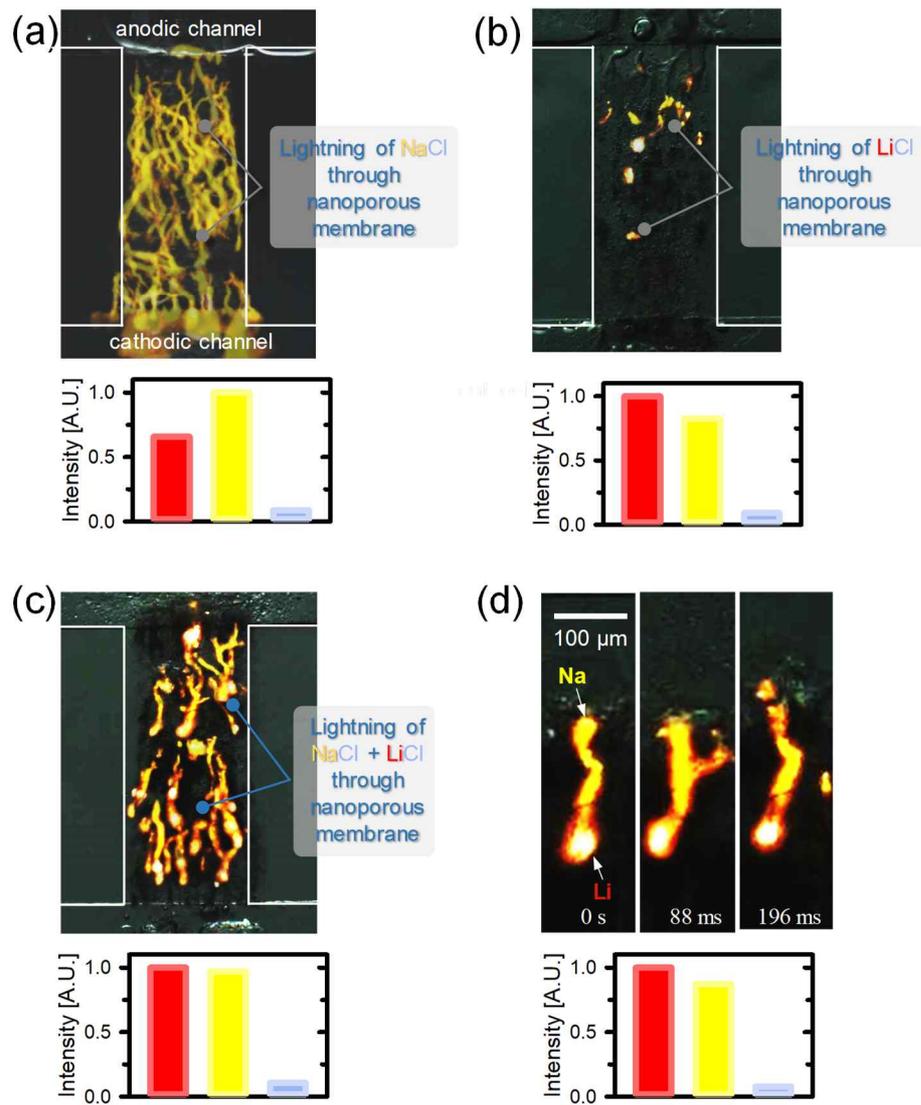

Figure 3